 \documentclass[final,5p,times,twocolumn]{elsarticle}
\usepackage{subfig}

\usepackage{amssymb}

\begin{document}

\begin{frontmatter}



\author[add1]{Bishal Chhetry}
\address[add1]{North Eastern Regional Institute of Science and Technology,
Nirjuli, Itanagar, Arunachal Pradesh 791109, India.}
\ead{bishalchhetry97@gmail.com}
\cortext[cor1]{corresponding Author}
 \author[add1]{Ningrinla Marchang}
 \ead{ningrinla@gmail.com; nm@nerist.ac.in}
 

\title{Detection Of Primary User Emulation Attack (PUEA) In Cognitive Radio Networks Using One-Class Classification}




\begin{abstract}
Opportunistic usage of spectrum owned by licensed (or primary) users is the cornerstone on which 
the Cognitive Radio technology is built. Unlicensed (or secondary) 
users that thus use the spectrum rely opportunistically on spectrum 
sensing to determine the presence of primary user signal. 
In such a context, an attacker may mimic the behavior of a primary user (PU) to deceive the secondary users (SUs) into 
believing that a PU signal is present whereas it is not. Such an attack is known as the 
Primary User Emulation Attack (PUEA). 
A malicious user may launch a PUEA with the intention of grabbing the vacant bands for its own transmission. Another reason may be to simply disrupt the functioning of the Cognitive Radio Network (CRN). 
This work investigates the use of 
one-class classification for detecting PUEA in an infrastructure-based CRN. We opine 
that sensing data collected at the fusion center mainly for Collaborative Spectrum Sensing (CSS) 
can be exploited to characterize a PU signal. The PU signal features thus learned can aid in 
distinguishing a PU signal from a PU signal emulation. In particular, we investigate the use of  
one-class classification techniques, viz., Isolation Forest, Support Vector Machines (SVM), Minimum Covariance Determinant(MCD) and  Local Outlier Factor(LOF) for detection of PUEA attacks. Simulation results 
support the validity of using one-class classification for detection of PUEA. 


\end{abstract}




\end{frontmatter}


\section{Introduction}
\label{}
The last few decades have witnessed the mushrooming of wireless technologies and services which has led 
to crowding the ISM bands, which will eventually lead to 
`\emph{radio traffic jam}'. Ironically, studies have shown that whereas some bands are overcrowded, some 
others, and a lot of the licensed bands in particular, lie fallow most of the time and in most areas. This 
has led to the emergence of cognitive radio technology. The basic idea of this technology is to make 
opportunistic use of these fallow bands   
without causing harmful interference to licensed users, which own these bands. These licensed users are also known 
as primary users (PUs). 
Thus in CRNs, unlicensed or secondary users (SUs) look out for unused licensed spectrum bands 
with the help of spectrum sensing. 
One important underlying principle upon which opportunistic usage 
in CRN works is that SUs must not cause harmful interference to PUs. This implies two things. 
First, 
only when an SU detects that a band is unused, then only, it can use the band 
for its transmission. Second, 
when an SU is 
transmitting in an unused band and the PU starts its transmission, the SU must vacant the band immediately. 

Unfortunately, such a requirement opens the door for attacks such as the 
as the Primary user emulation (PUE) attack. This attack happens when 
an attacker (which could be an SU acting maliciously) emulates a PU signal. 
Consequently, if other honest SUs  
happen to be using the associated channel, they will evacuate from the channel. On the other 
hand, if they happen to be waiting for this channel to be free, they will not be able to use it as they 
would mistakenly sense that PU signal is present. An inside attacker, viz., a 
malicious SU,  may launch a PUEA for grabbing 
the channel selfishly for itself. On the other hand, an outside attacker may do so  
for causing a DoS (Denial of Service) attack. 
Hence, the PUEA greatly limits the spectrum access opportunity of SUs.

Whereas there are several PUEA mitigation techniques in the literature, unlike them, 
this work proposes the 
use of one-class classification for the same. The basic motivation behind this work is: `The sensing data 
collected at the fusion center (FC) forms a rich audit data-set for understanding the features of a 
PU Transmission. One-class classification is a perfect fit, as it can use this imbalanced 
data-set for training a model. 
The trained model can be used for detecting any PU-like transmission which do not fit into this class 
of \emph{`PU transmissions.'} 

The contributions of this paper can be summarized as follows:
\begin{itemize}\item One-class classification is used for detecting PUEA. Four classification algorithms are investigated, viz., Isolation Forest, Support Vector
   Machine, MCD and LOF.
\item Audit data of the sensing reports collected at the FC are processed to characterize the behaviour 
of a PU transmission, thereby generating the training data-set. 
\item Simulation results support the validity of using one-class classification for PUEA detection.
\end{itemize}
 
The rest of this paper is structured as follows. Section \ref{related} reviews related existing works, which is followed by the system model in section \ref{sysmodel}. Then, we explain the development of the proposed scheme in section \ref{detection}. In section \ref{result}, we present the  results and performance analysis which is followed by the conclusions in section \ref{conclusion}.

\section{Related Work}\label{related}
Research on detecting PUE attacks is an ongoing activity. 
A lot of work in this area 
has been reported in the literature \cite{Anand} - \cite{Furqan}. 
In \cite{Anand}, an analytical  model  and  a practical mechanism to 
detect PUEA in  CRNs (a.k.a. dynamic  spectrum  access networks) are presented. In a 
similar work \cite{Jin}, 
an analytical approach using Fenton’s approximation and Markov inequality to 
obtain a lower bound on the probability of a successful PUEA is presented, in which 
a fading wireless environment is considered.
In \cite{Baldini}, we find a survey on security threats and related protection 
techniques for Software Defined Radio (SDR) and Cognitive Radio (CR) technologies. 
Estimation techniques (e.g., maximum likelihood estimation) and the learning methods (e.g., mean-field approach) can also be used for mitigating PUEA, wherein 
individual spectrum decision is obtained by characterizing the received power 
at good secondary user through a flexible log-normal sum approximation method 
\cite{Zesheng}. 

An example of a technique based on localization is presented in \cite{Chen}, in which 
a transmitter verification scheme, called LocDef (localization-based defense) is 
given which verifies whether a given signal is that of an 
incumbent transmitter by estimating its location and observing its 
signal characteristics. The scheme presented in 
\cite{Xin} uses  the activity pattern of a signal through spectrum sensing, 
such as the ON and OFF periods of the signal. Then, the observed signal activity 
pattern is reconstructed through a reconstruction model. The reconstruction error is used 
to distinguish a signal activity pattern of a PU from a signal activity pattern of a PUEA 
attacker. A comparatively involved scheme \cite{Lui} uses cryptographic signatures 
and wireless link signatures (derived from physical radio channel characteristics) to enable primary user detection in the presence of PUEA attackers.
In \cite{Julio}, a framework for security and resilience 
is applied to spectrum sensing functionality to ensure that it is robust even 
in the event of failures and PUEA.

In \cite{Morgan}, the authors develop a distributed spectrum decision protocol in which SUs make individual spectrum decisions and then exchange individual sensing results with their one-hop neighbors to increase resilience to PUEA. 
In \cite{Sureka}, a mechanism for detection and prevention of PUEA 
is presented, which employs Time-Distance with signal Strength Evaluation (TDSE) and Extreme Machine Learning (EML) algorithm. Whereas our work also uses machine learning, it 
relies only on the audit data of sensing reports received at the FC. However, 
in 
\cite{Sureka}, the authors use time duration of traversal, RSS from the oncoming transmitter and 
the ideal time of the PU. 
A novel algorithm for detecting non-intelligent primary user emulation attack 
is presented in \cite{Yuan}. 
Neyman-Pearson composite hypothesis test (NPCHT) and Wald’s sequential probability 
ratio test (WSPRT) are employed in 
\cite{Zjin}  
to detect PUEA. In \citep{Zhang}, the authors describe the special characteristics 
of cognitive radio and CRN and present an analysis the potential security threats that 
arise  due  to  these characteristics. 
A method to detect the  PUE attack of  mobile PUs by exploiting the correlations between RF signals and  acoustic information with the aim of verifying the existence of 
wireless microphones is found in 
\citep{Zeng}. 
In \citep{Subba}, a spectrum  decision  protocol  resilient to PUEA in dynamic spectrum 
access networks is proposed.
Device-specific features have been employed to propose a passive, non-parametric classification method to determine the number of PUs that are transmitting in the spectrum.
In \cite{Furqan}, an algorithm for the detection of PUEA and jamming attacks in CRN 
is proposed, which is founded on the sparse coding of the compressed received signal 
over a channel-dependent dictionary. 


Our work is motivated by the above tools and techniques. However, none of them have used one-class classification which we perceive 
is a suitable technique for detecting PUEA. 
Moreover, just by analyzing the sensing reports received at the FC, a PU transmission can be characterised and the derived features are used to drive the classification model. 
To the best of our knowledge, our work is the first one which employs sensed data at the FC to train one-class classification 
models for detecting PUEA. Representative algorithms such as 
Isolation Forest (IF), SVM, MCD and LOF are employed in our work. 

\section{System Model} \label{sysmodel}
\begin{figure}[h]
\centering
\includegraphics[height=2in]{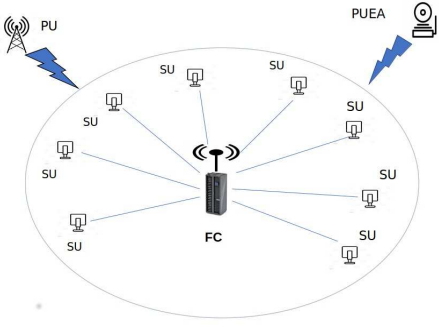}
\caption{PUEA in an Infrastructure-based CRN}
\label{figureSystemModel}
\end{figure}

We assume a CRN consisting of $n$ SUs, wherein all the SUs collaborate to 
sense the presence of a PU. 
We assume that the PU and the SUs are stationary, and the PU is away from the SU. 
No two SUs have the same location. 
The system model is illustrated in Fig. \ref{figureSystemModel}. 

For simplicity, we consider a single-channel system. 
Each SU takes the help of an energy detector for sensing the channel locally \cite{Zesheng} and sends its sensing report to the FC. 
Here, we assume a continuous reporting CRN and so, each SU sends the energy 
level (in dB) of the received signal to the FC. The FC uses some fusion rule to make 
the final decision about whether the PU is present or not. However, how 
the decision is made is not the focus of our study. Rather, it focuses on exploiting these sensing reports (raw energy levels) received  
at the FC for mitigating PUEA. 

Time for each SU is divided into fixed-size slots and these slots are synchronized. 
Each slot is again subdivided into two sub-slots: i) sensing, and ii) data transmission, wherein the first sub-slot is for sensing and the second one for data transmission.  
In each sensing sub-slot, an SU senses the channel and sends the sensing report to the FC. 
Thus, in this sub-slot, the FC receives the reports (energy values), $y_1, y_2,..,y_n$ 
from the $n$ SUs. 

The relationship between the transmit signal power and the received 
signal power is modelled with the help of the  
path loss model and the log-normal shadowing of a communication 
channel as in   
\cite{Zesheng}. 
Assuming the path loss exponent,  $\alpha$, the received signal energy $P_{r}$ 
is proportional to r\textsuperscript{$-$$\alpha$}, where $r$ is the distance between 
the transmitter and the receiver. Additionally, $P_{r}$ is also proportional to a shadowing random variable G. Here, G=$\ e^{a \beta}$, where $a=\frac{ln10} {10}$ and $\beta$ follows a normal distribution, $N(0,\sigma^2$). 
Thus, when a transmitter employs energy $P_{t}$ to transmit signals, then the received signal energy, $P_{r}$ at the receiver end is given by \cite{Zesheng}: 
\begin{eqnarray}
P_{r} = P_{t} r\textsuperscript{-$\alpha$} e\textsuperscript{a$\beta$} \label{eq:received}
\end{eqnarray}

\par 
At the FC, these received energy values are processed to create the data-set, which 
is described later. The built data-set is used for one-class classification with the 
goal of mitigating PUEA. 
As seen in Fig. \ref{figureSystemModel}, a PUEA can be launched by an outsider with the 
aim of disrupting the functioning of the CRN. Alternatively, it can also be launched 
by an insider, i.e., one among the SUs. 

 
\section{PUEA Detection as a One-Class Classification Problem}\label{detection} 
This section describes how we employ one-class classification for detecting PUEA attack in a CRN. First we describe  how the detection problem can be transformed into a one-class classification problem. Second, we present how data-sets are formed and  can be used for analysis. Third, we explain in brief the classification algorithms, viz., Isolation Forest, SVM, MCD and LOF that have been used in our work.

\subsection{One-Class Classification}
In classification, it may happen that  
the training 
set is such that all or most of the examples in the set are from one class 
(say, the positive class), and none or very little from the negative class. Such a classification 
is called one-class classification. After the model is
 trained using such a data-set, a new example which is fed to the trained model would 
 be classified into either the positive class of the negative class. When the 
 training set consists of only examples from one class, an input example which does 
 not get classified into this class is also called an outlier. Hence, one-class 
 classification techniques are also known as outlier detection techniques. 

\subsection{Motivation}
Any transmission which imitates a PU transmission falls under the gambit of PUEA. 
Since the only requirement of this attack is that normal SUs should unwittingly sense it 
as a PU signal, there could be several ways this attack is carried out. For 
instance, an attacker could position itself as close as possible to a PU station and 
launch an attack. Alternatively, it could emulate the PU signal (e.g., PU's 
transmission power). Thus, it is a challenging task to profile a PUEA behaviour. On 
the other hand, profiling of a PU behaviour is easier comparatively, since a PU does 
not change its form over long periods of time. For instance, most of the PU stations 
are stationary stations (e.g., a TV tower) which do not generally change its 
transmission characteristics over long periods of time. In conclusion, the idea is 
to build a training data-set that consists of only PU signal examples. Consequently, 
one-class classification is a perfect fit for detecting PUEA. 

\subsection{Building the Data-set}\label{BuildingDataset}
At each slot, when the FC receives the sensing reports (energy levels) from the $n$ 
SUs, it collects them and calculates the mean, variance, median, upper quartile and 
lower quartile of the $n$ values and creates a training example. 
Thus, each training example consists of five input features: mean, variance, median, upper quartile, lower quartile. 
We include output feature (i.e., label) as `1' (positive class)    
when PU is present. On the other hand, if the example is from a PUEA, it would be 
labelled `-1' (negative class). 
So, if we collect data from $T$ slots to create the data-set, there would be $T$ 
examples. in which PU is present, there will be $T$ examples. 


\subsection{Classification Algorithms}
The  data gathered at the FC for a good PU  is processed so as 
to train a classifier. Subsequently, we can employ the trained classifier to 
predict whether the signal is from a PU or an attacker. We use four classifiers, 
viz., Isolation Forest, Support Vector Machine (SVM), 
Minimum Covariance Determinant (MCD) and 
Local Outlier Factor (LOF). 

\subsubsection{Isolation Forest}\label{Isolation Forest}
 Isolation Forest \cite{IF} 
 is a tree-based anomaly detection algorithm. 
Anomalies are 
isolated with the help of tree structures. The trees are created in such a way 
that isolated examples tend to have a relatively short depth in the trees as 
compared to normal examples which are less isolated, and thereby 
exhibit a greater depth in the trees. 

\subsubsection{Support Vector Machine (SVM)}\label{Support Vector Machine}
Initially, the SVM algorithm was developed for binary classification. However, 
it can be employed for one-class classification \cite{SVM}. 
For SVM to behave as a one-class classification technique, the algorithm works by capturing the density of the majority class. Then, 
examples on the extremes of the density function are classified as outliers. The SVM thus modified is known as One-Class SVM. 

\subsubsection{Minimum Covariance Determinant (MCD)}\label{Minimum Covariance Determinant}
MCD \cite{MCD} algorithm works by creating a 
a hypersphere (ellipsoid) in such a way that it covers the normal data (examples). Hence, any example that is not covered by this 
shape is detected as an outlier. MCD is an efficient implementation of 
this technique for multivariate data. 


\begin{figure}[h]
\centering
\subfloat[Inside SU Region]{
\includegraphics[width=0.48\columnwidth, height=1.5in]{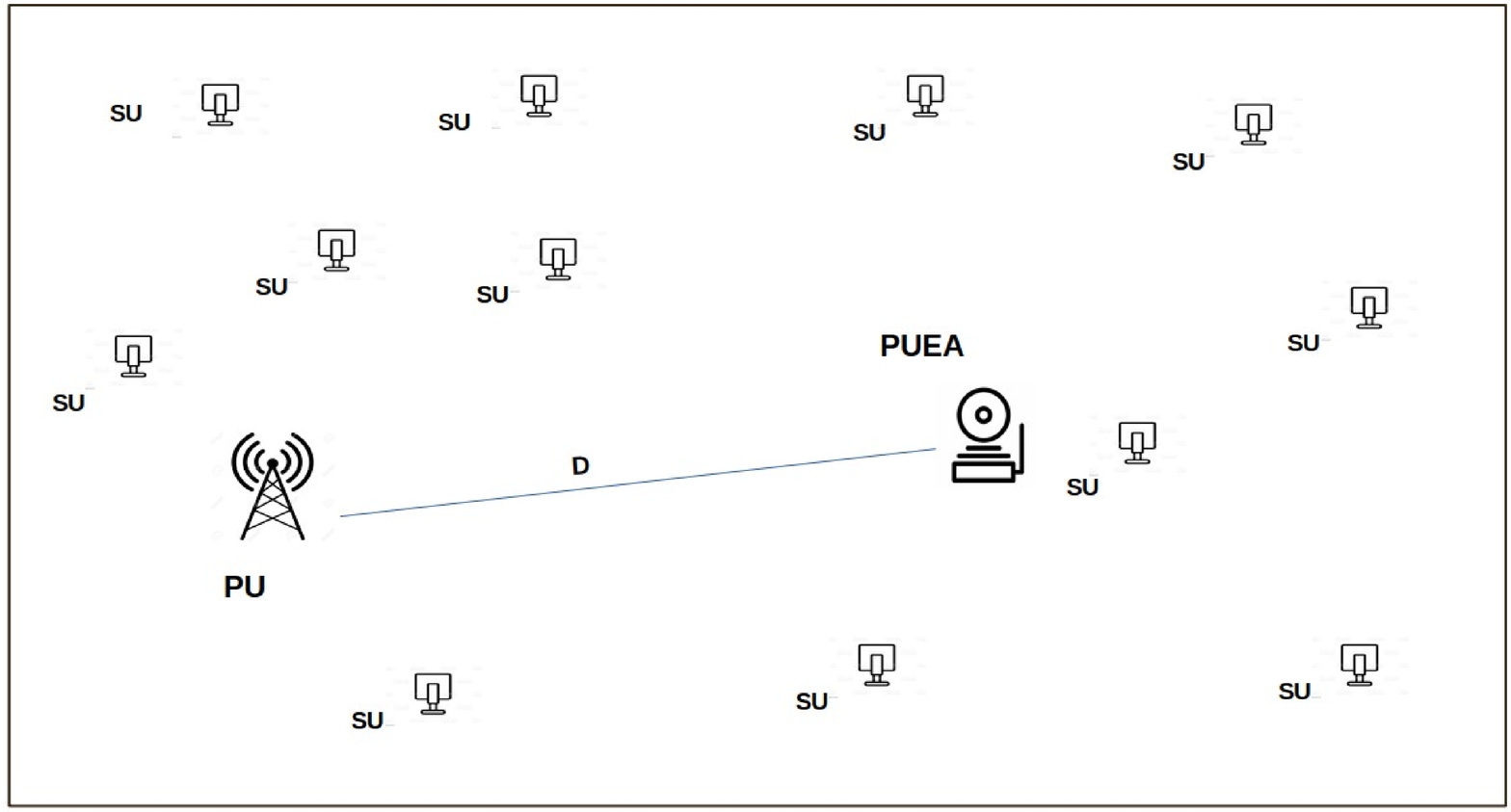}}
\subfloat[Outside SU Region]{
\includegraphics[width=0.48\columnwidth, height=1.5in]{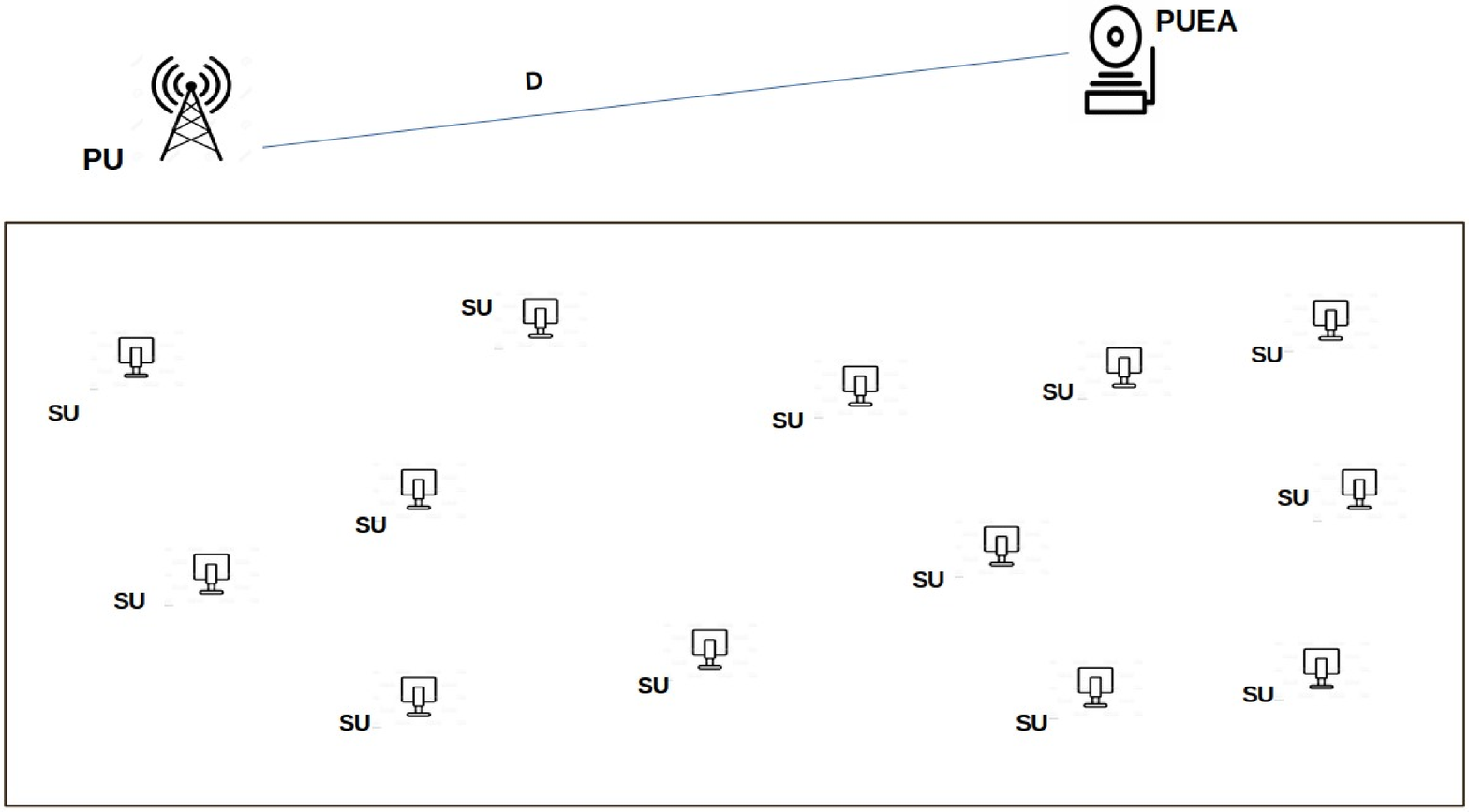}}
\caption{Locations of PU and Attacker}
\label{figureSUregion}
\end{figure}


\subsubsection{Local Outlier Factor(LOF)}\label{Local Outlier Factor (LOF)}
Another approach \cite{LOF} to detect outliers is to identify those examples 
that are far from the other examples, which form larger groups in the feature space. This approach works by determining the LOF degree of each example with the 
help of nearest neighbors concept. 
This technique is known to be more reliable for less number of features as 
compared to when the 
number of features is high. 

\begin{figure}[h]
\centering
\subfloat[$10\%$ PUEA data]{
\includegraphics[width=0.48\columnwidth, height=1.5in]
{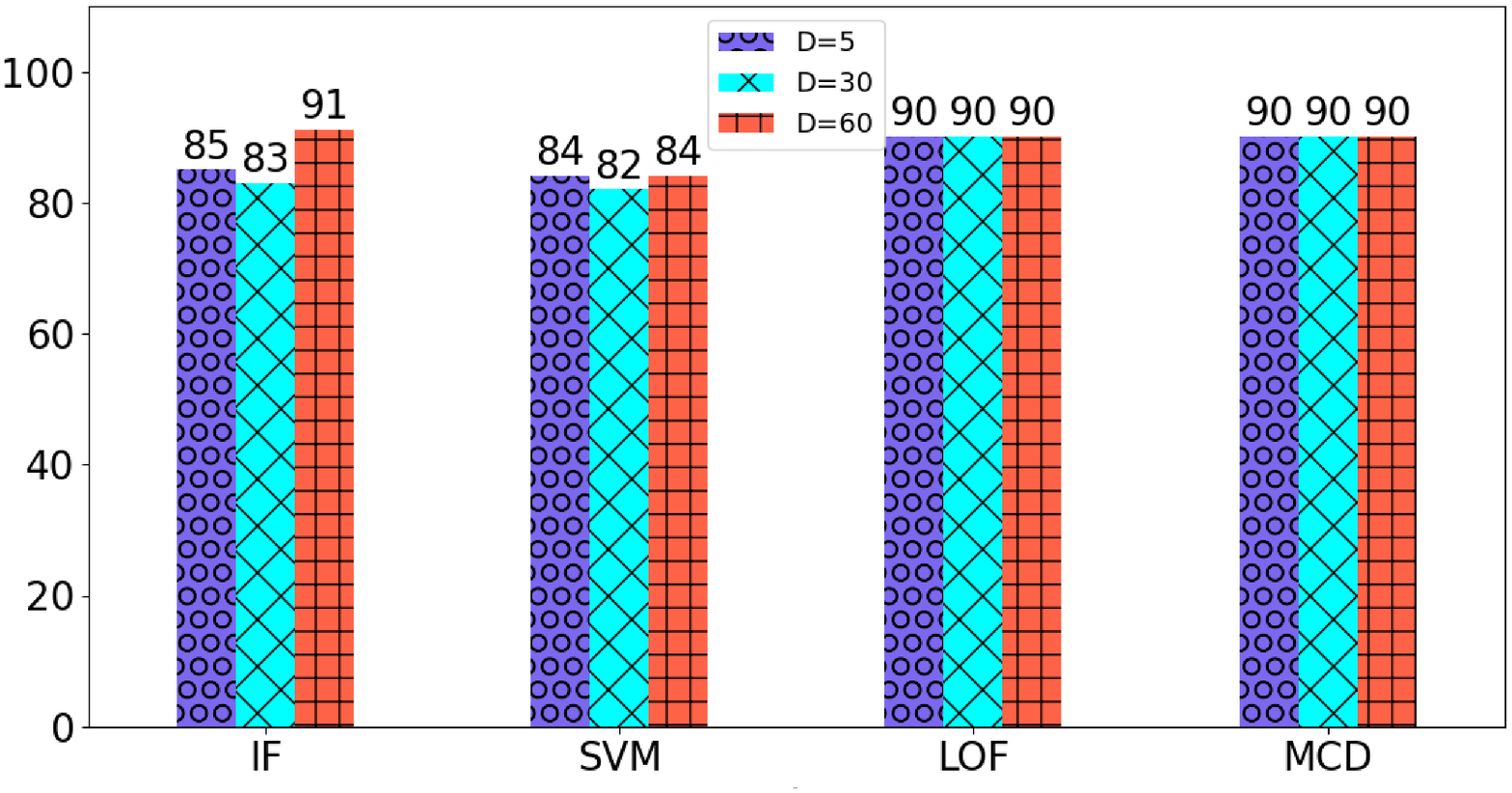}}
\subfloat[$20\%$ PUEA data]{
\includegraphics[width=0.48\columnwidth, height=1.5in]
{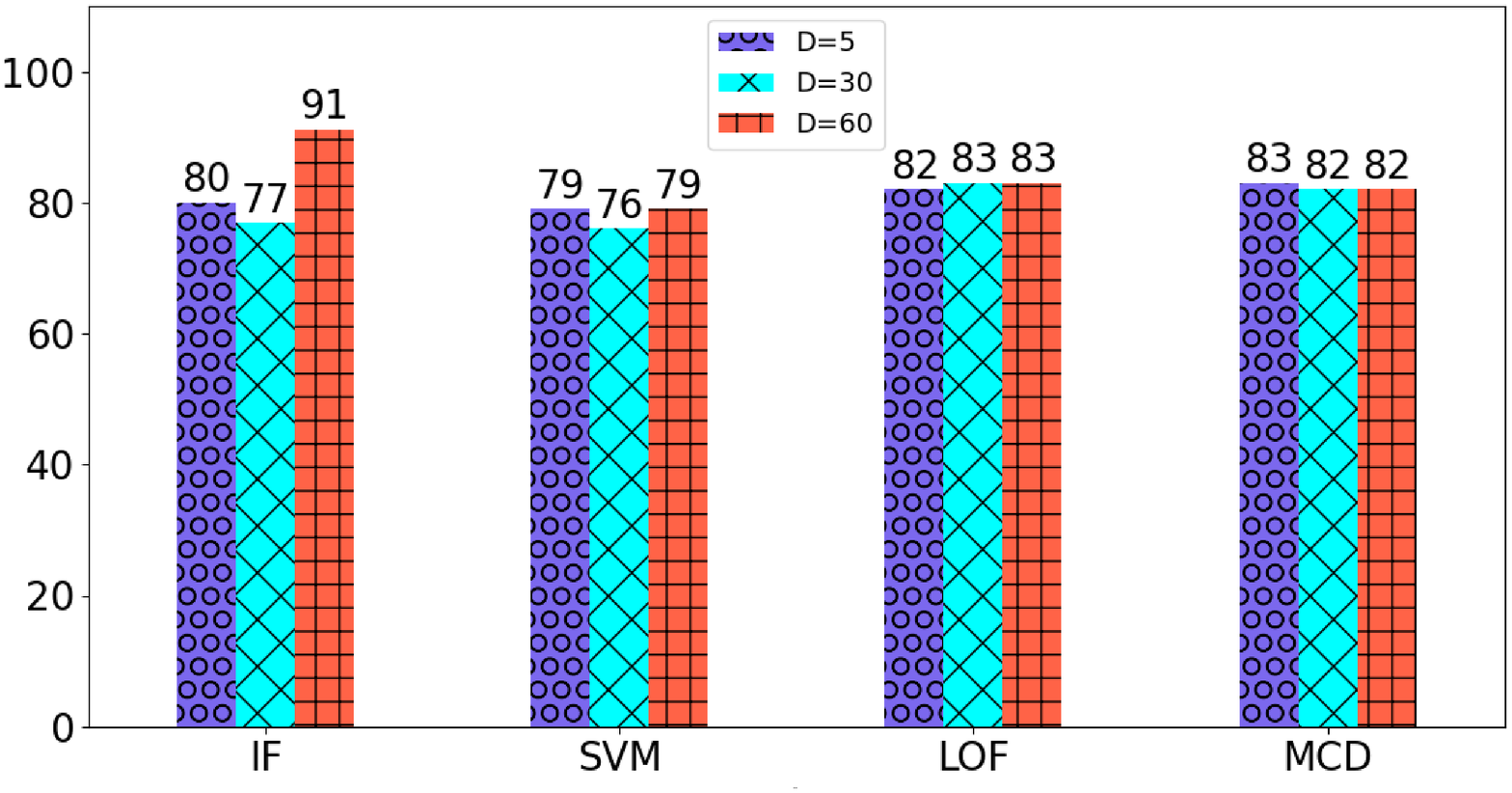}}
\caption{Accuracy: [Inside SU region]}
\label{figureAccuracyInside}
\end{figure}

\begin{figure}[h]
\centering
\subfloat[$10\%$ PUEA data]{
\includegraphics[width=0.48\columnwidth, height=1.5in]
{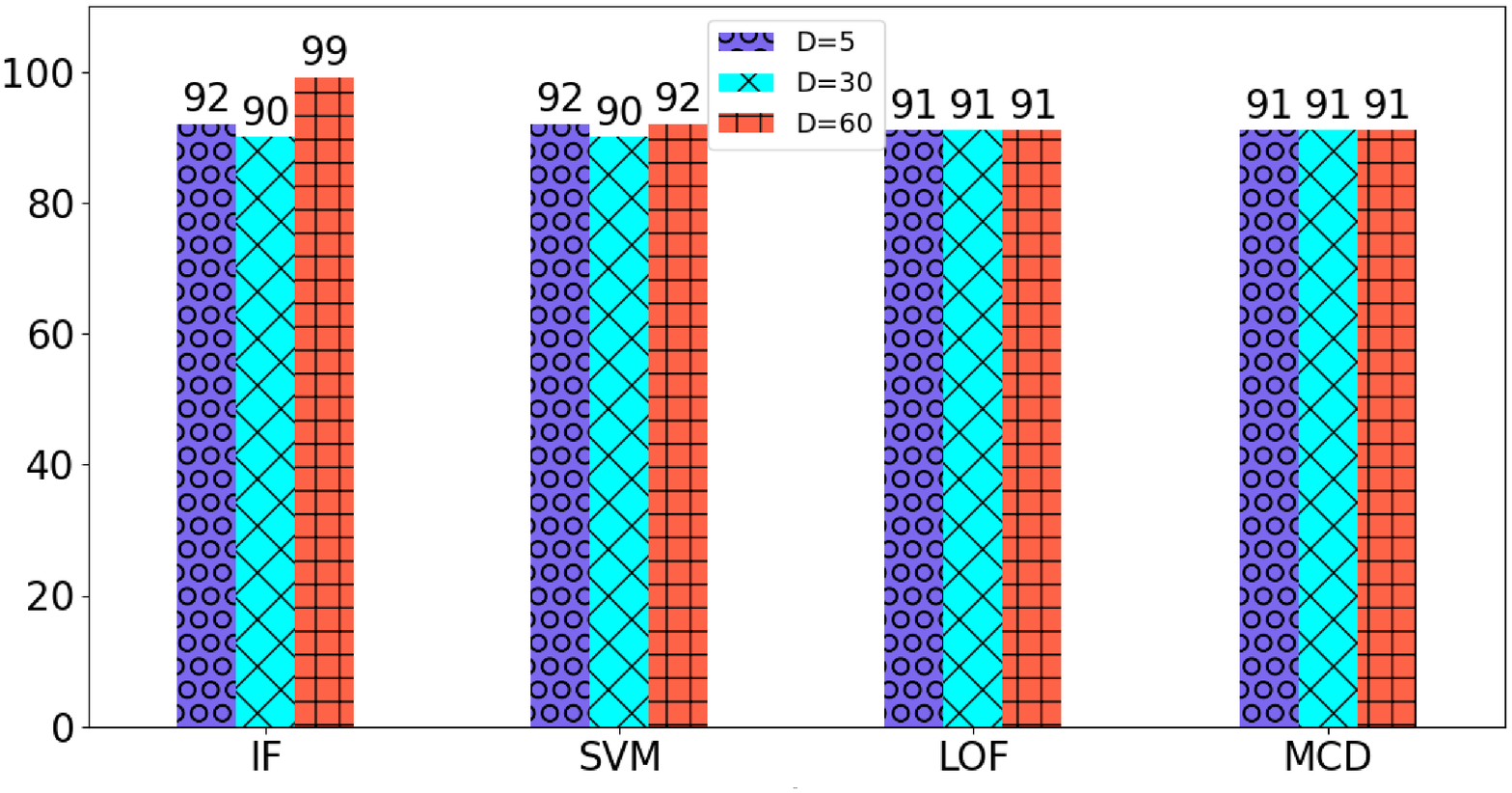}}
\subfloat[$20\%$ PUEA data]{
\includegraphics[width=0.48\columnwidth, height=1.5in]
{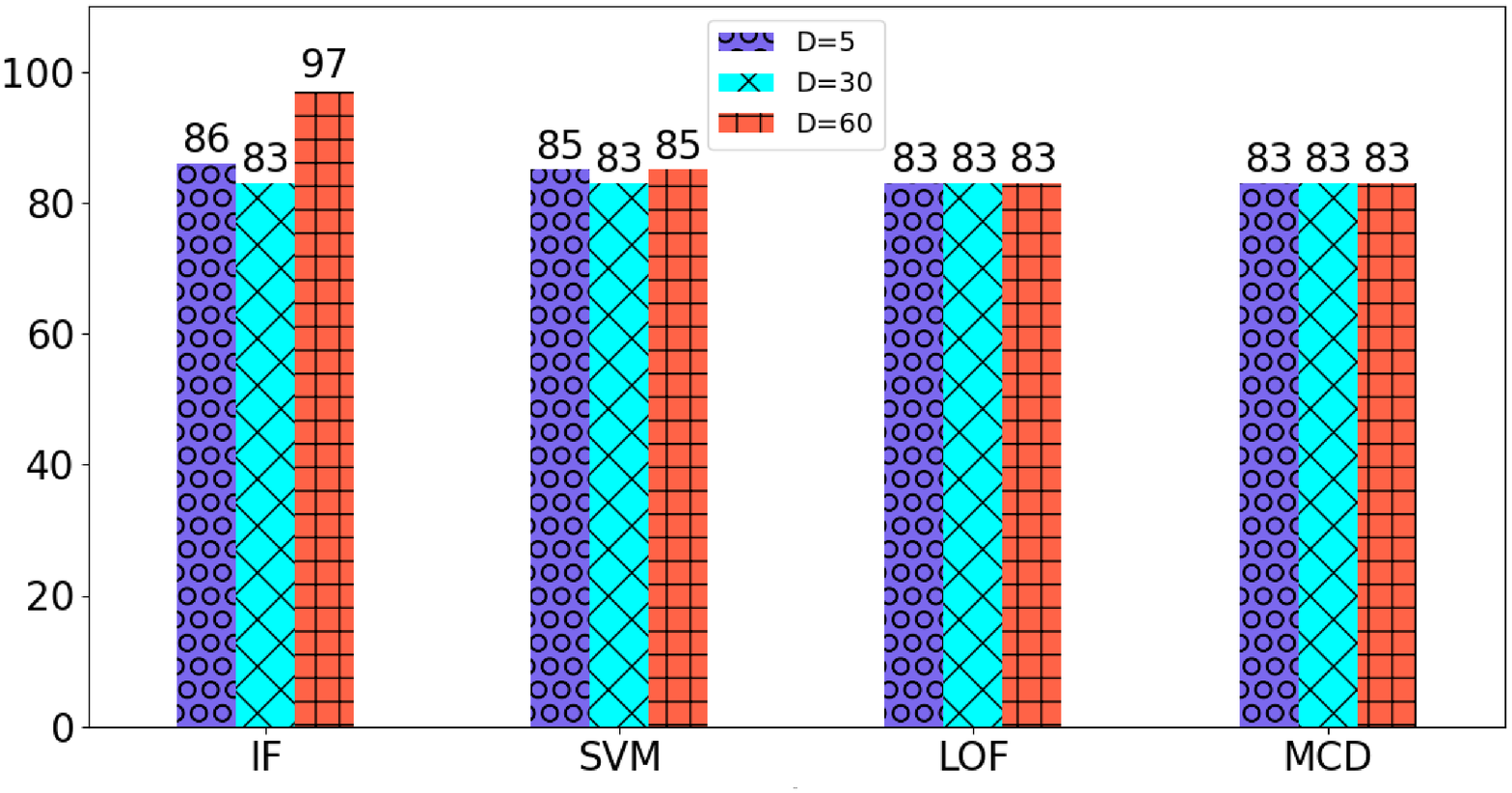}}
\caption{Precision: [Inside SU region]}
\label{figurePrecisionInside}
\end{figure}

\begin{figure}[h]
\centering
\subfloat[$10\%$ PUEA data]{
\includegraphics[width=0.48\columnwidth, height=1.5in]
{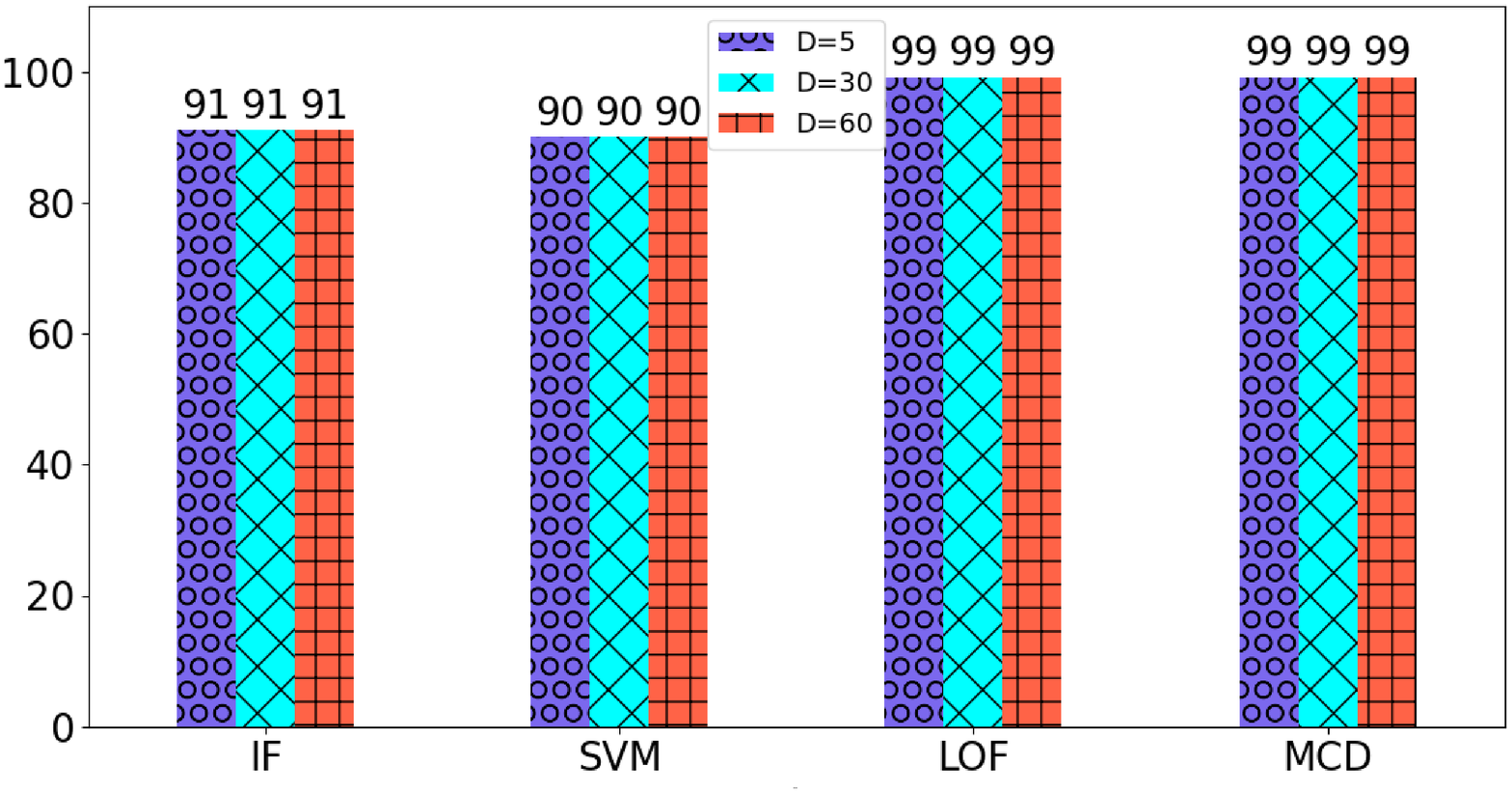}}
\subfloat[$20\%$ PUEA data]{
\includegraphics[width=0.48\columnwidth, height=1.5in]
{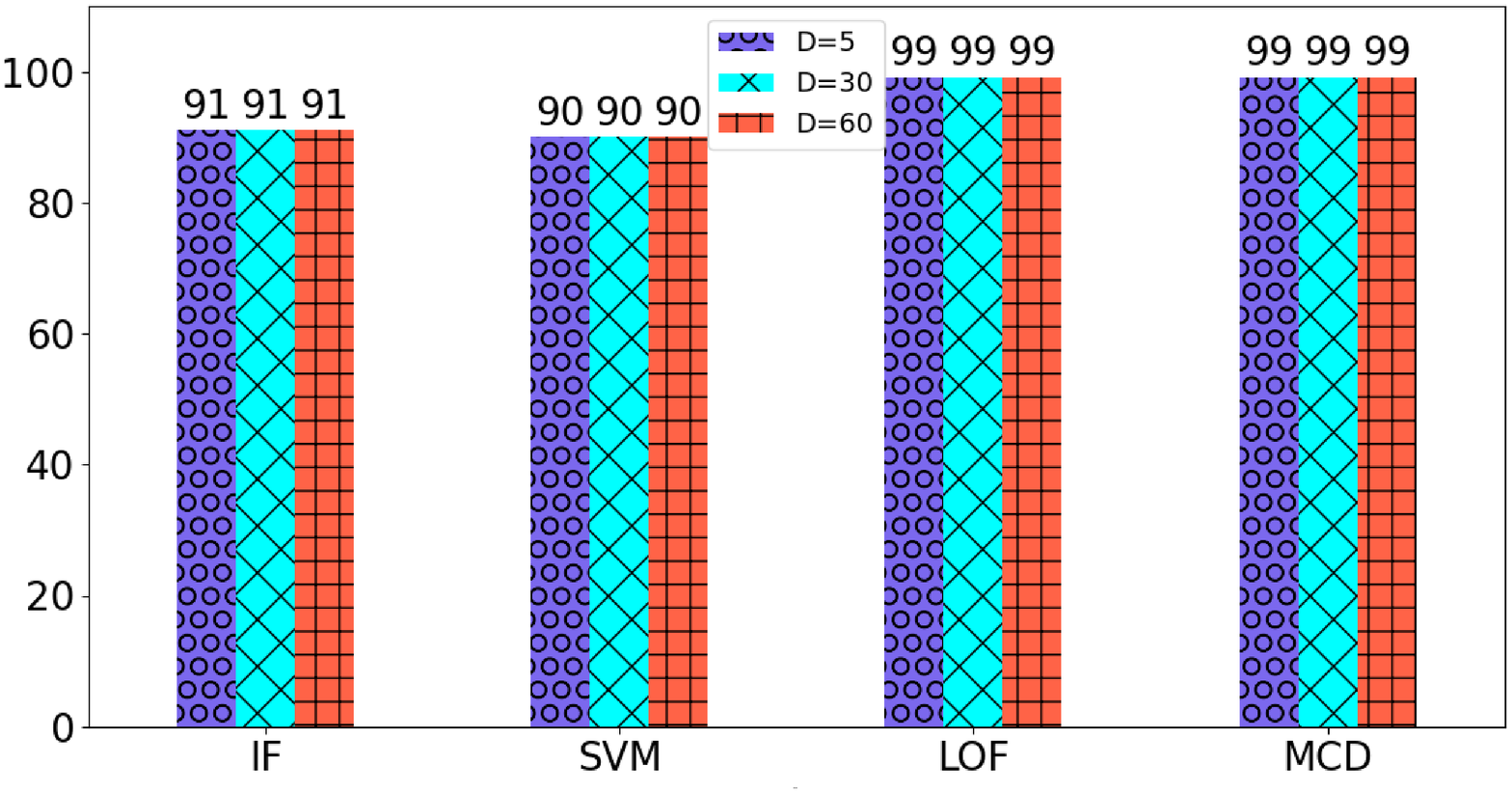}}
\caption{Recall: [Inside SU region]}
\label{figureRecallInside}
\end{figure}

\begin{figure}[h]
\centering
\subfloat[$10\%$ PUEA data]{
\includegraphics[width=0.48\columnwidth, height=1.5in]
{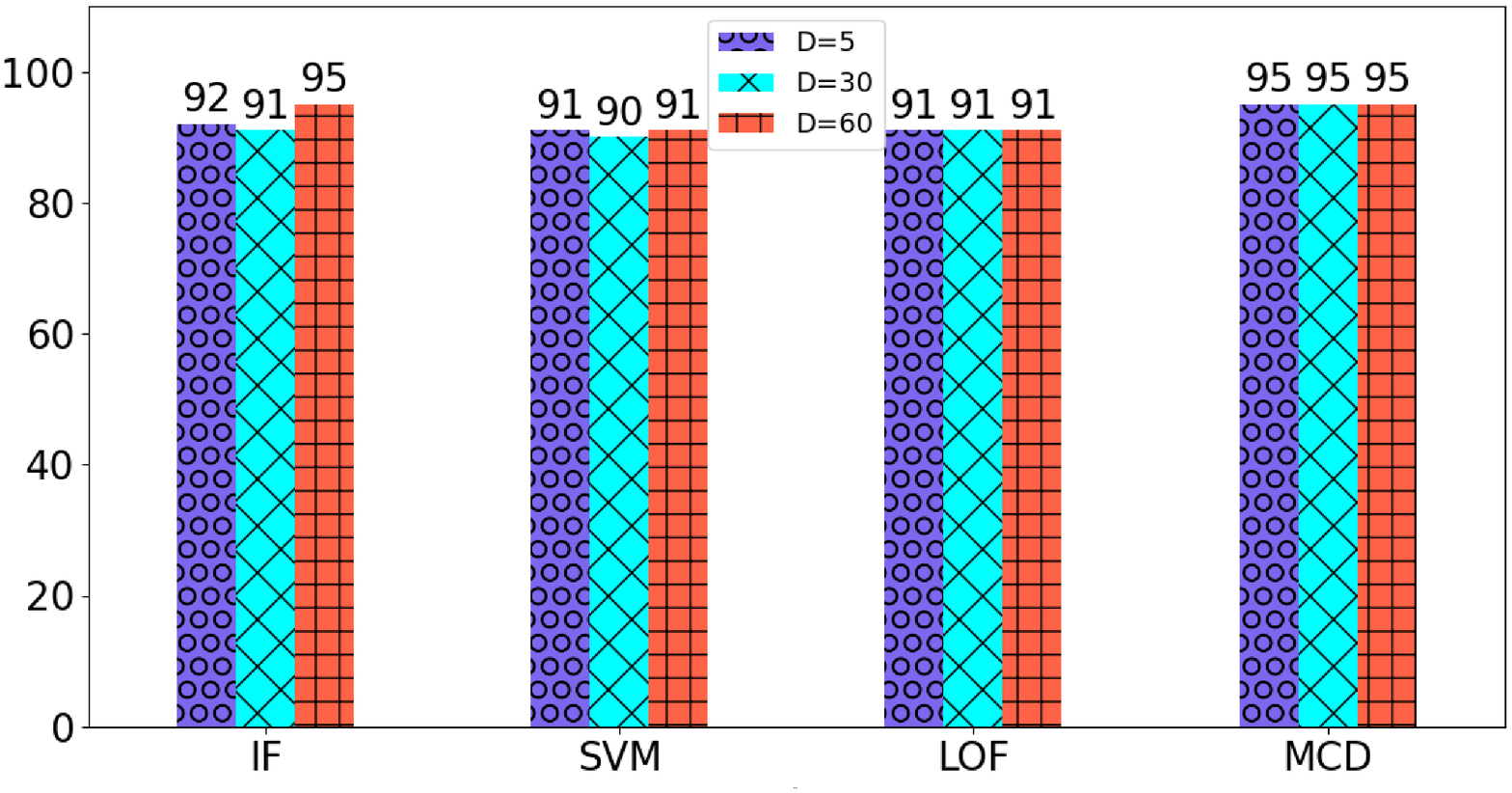}}
\subfloat[$20\%$ PUEA data]{
\includegraphics[width=0.48\columnwidth, height=1.5in]
{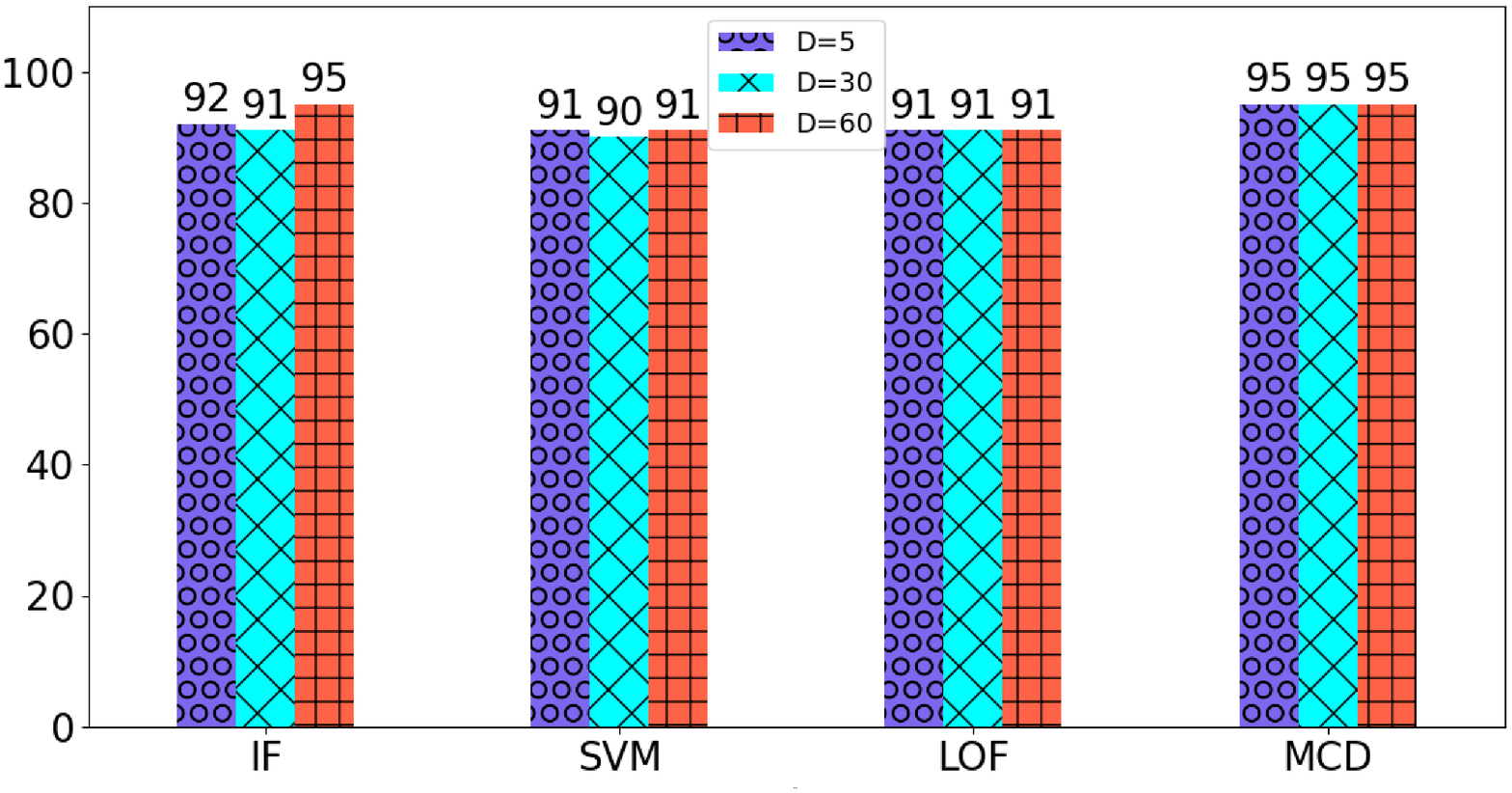}}
\caption{F1-score: [Inside SU region]}
\label{figureF1scoreInside}
\end{figure}

\section{Results and Discussion}\label{result}
This section presents the simulation results. We
consider a CRN consisting of 40 SUs and 10,000 sensing slots. 
We assume that the SUs occupy some region of $100\times100$ sq. units. Then, we randomly generate the locations of the SUs 
in this region. Next, we choose a location each for the PU and 
the attacker. Depending on how their locations are chosen, we consider two scenarios: 
(i) when the PU and the attacker are within the same region as the SUs as 
depicted in Fig. \ref{figureSUregion}(a), and (ii) when the PU and the attacker are 
outside the SUs' region (refer Fig. \ref{figureSUregion}(b)). In both the figures, 
$D$ denotes the distance between the PU and the attacker. 

The model described in Section \ref{sysmodel} is used to create the 
data-set. 
We assume the same paramater values used in 
\cite{Zesheng}. Thus, we let  
$P_{t} = 10$ and  
$\sigma^2 = 8$ when PU is transmitting. Whereas, 
when the attacker is transmitting, the channel parameter is $\sigma^2$ = 4, 8, or 12.

Using the above model, the energy levels received 
from the PU at the SUs are generated for each slot. Then, they are processed to 
generate the examples, each example consisting of 
input features, viz., 
the mean, variance, median, upper quartile and 
lower quartile of for each slot. 
Here, we assume that a common control channel is used to send the 
energy levels from the SUs to the FC and that there is no channel error. 
Hence, the data-set is created at the FC. 
Since 10,000 slots are considered, the 
size of the training data-set is 
10,000. 

Similarly, for testing purpose, the energy levels 
that are received at the SUs from the attacker are artificially generated using the 
same model. Due to lack of availability of real-world 
data-set w.r.t. PUEA, we build an artificial data-set as 
explained above. 

We say that a true positive has occurred if a malicious PUE attack is 
detected as malicious, whereas a false positive has occurred if a geniune 
PU is incorrectly detected as an attacker. Similarly, we say that a true negative has occurred if 
a good PU is detected as a genuine PU, whereas a false negative has occured 
if an attacker is identified as a genuine PU. 

Let TP, TN, FP, FN and N denote the count of true positives, true negatives, 
false positives, false negatives and the total number of examples. 
Note that $N=TP+TN+FP+FN$. 
The performance parameters considered are: 
i) $Accuracy$: It gives a measure of how many cases are correctly 
identified. 
It is defined by term, $(TP+TN)/N$. 
It is most useful when all the classes are equally important.
ii) $Precision$: It gives a measure of how many out of 
all the predicted positive cases are correctly identified as positive cases. 
It is defined by term, $TP/(TP+FP)$. 
Thus, this value will be 
high when FP is less. 
iii) $Recall$: It gives a measure of how many out of 
all the actual positive cases are 
correctly identified as positive cases.
It is defined by term, $TP/(TP+FN)$. 
Thus, this value will be 
high when FN is less. 
iv) $F1-Score$: It is the harmonic mean of Precision and Recall. Thus, this 
value will be high when the 
count of incorrectly classified cases is less. 
The values of these metrics shown in the following result graphs are in terms of percentage. 

The simulation results are divided into two parts based on the location of the PU and the attacker 
w.r.t. the region where the SUs lie. We consider two scenarios as shown in Fig. \ref{figureSUregion}. 
This helps us to determine whether such factors have any 
impact on the detection system.

\subsection{PU Inside SU Region}
In the scenario when the PU and the attackers are within the region where the SUs lie (refer Fig. \ref{figureSUregion}(a)), we compare the results for varying distances between the PU and the attacker. Moreover, we vary the percentage of PUEA examples in the test set and determine the results. 
It may be noted here that the training set consists of only examples 
associated with the PU signal (i.e., 10,000 examples) as explained in section 
\ref{BuildingDataset}. However, with regard to the test set, we initially form it using the 
training set (consisting of 10,000 examples) and PUEA examples whose count is $10\%$ of the 
size of the training set. Thus, the test set consists of 11,000 examples. Next, we increase 
the percentage of the PUEA examples in the test set from $10\%$ to $20\%$. The accuracy shown by different algorithms for varying values of `D' 
are shown in 
Fig. \ref{figureAccuracyInside}.  
The label `D=5' indicates the plot when the distance between the PU and the attacker is 5 units. 
Similar labels in the rest of the graphs have similar meanings. 
In Fig. 
\ref{figureAccuracyInside}(a), we find that LOF and MCD outperform the other algorithms 
with an accuracy of $90\%$. We find that the accuracy drops a little when the percentage 
of PUEA examples is $20\%$ (refer 
Fig. \ref{figureAccuracyInside}(b)) 
as compared to when it is $10\%$ (refer 
Fig. \ref{figureAccuracyInside}(a)). This is due to the increase in the number of unseen examples 
which are not in the training set. 
We also find that there is no perceivable 
difference between the results when the 
values of $D$ are varied.

\begin{table}[h!]
  \begin{center}    
    \begin{tabular}{|l|c|c|c|c|} 
\hline

\multicolumn{1}{|c|}{k} & IF & SVM & MCD & LOF\\
\multicolumn{1}{|c|}{ } & acc.(\%) & acc.(\%) & acc.(\%) &acc.(\%)\\
\hline
2 & 84 &84 &90&90\\
5& 83&84&90&90\\
10&84&84&90&90\\
20&84&85&90&90\\

\hline
  \end{tabular}
\caption{k-fold cross validation: 10\% PUEA data [Inside SU region]}
\label{table:inside10}
  \end{center}
\end{table}
\begin{table}[h!]
  \begin{center}    
    \begin{tabular}{|l|c|c|c|c|} 
\hline
\multicolumn{1}{|c|}{k} & IF & SVM & MCD & LOF\\
\multicolumn{1}{|c|}{ } & acc.(\%) & acc.(\%) & acc.(\%) &acc.(\%)\\
\hline
2 & 77 &79 &83&82\\
5& 77&79&82&82\\
10&77&79&82&82\\
20&76&78&82&82\\
\hline
  \end{tabular}
\caption{k-fold cross validation: 20\% PUEA data [Inside SU region]}
\label{table:inside20}
  \end{center}
\end{table}

When the data-sets are imbalanced as in our study, accuracy may not be a good parameter to measure 
performance. 
One important metric is precision. 
Fig. 
\ref{figurePrecisionInside} illustrates the corresponding precision values for the same 
scenarios. 
Precision is recorded to be above $90\%$ when the percentage of PUEA data in the 
test set is $10\%$ (refer Fig. \ref{figurePrecisionInside}(a)) whereas it is above $80\%$ when 
the corresponding percentage is $20\%$ (refer 
Fig. \ref{figurePrecisionInside}(b)) due to the reason 
stated earlier that there are more unseen 
examples in this case as compared to the 
earlier case, viz., $10\%$ case.

Recall is illustrated in Fig. \ref{figureRecallInside}. For both the scenarios (graphs), recall is 
above $90\%$ for all the algorithms. Moreover, the F1-score is also well above $90\%$ 
for all the algorithms for both scenarios 
(refer Fig. \ref{figureF1scoreInside}). 

\begin{figure}[h]
\centering
\subfloat[$10\%$ PUEA data]{
\includegraphics[width=0.48\columnwidth, height=1.5in]
{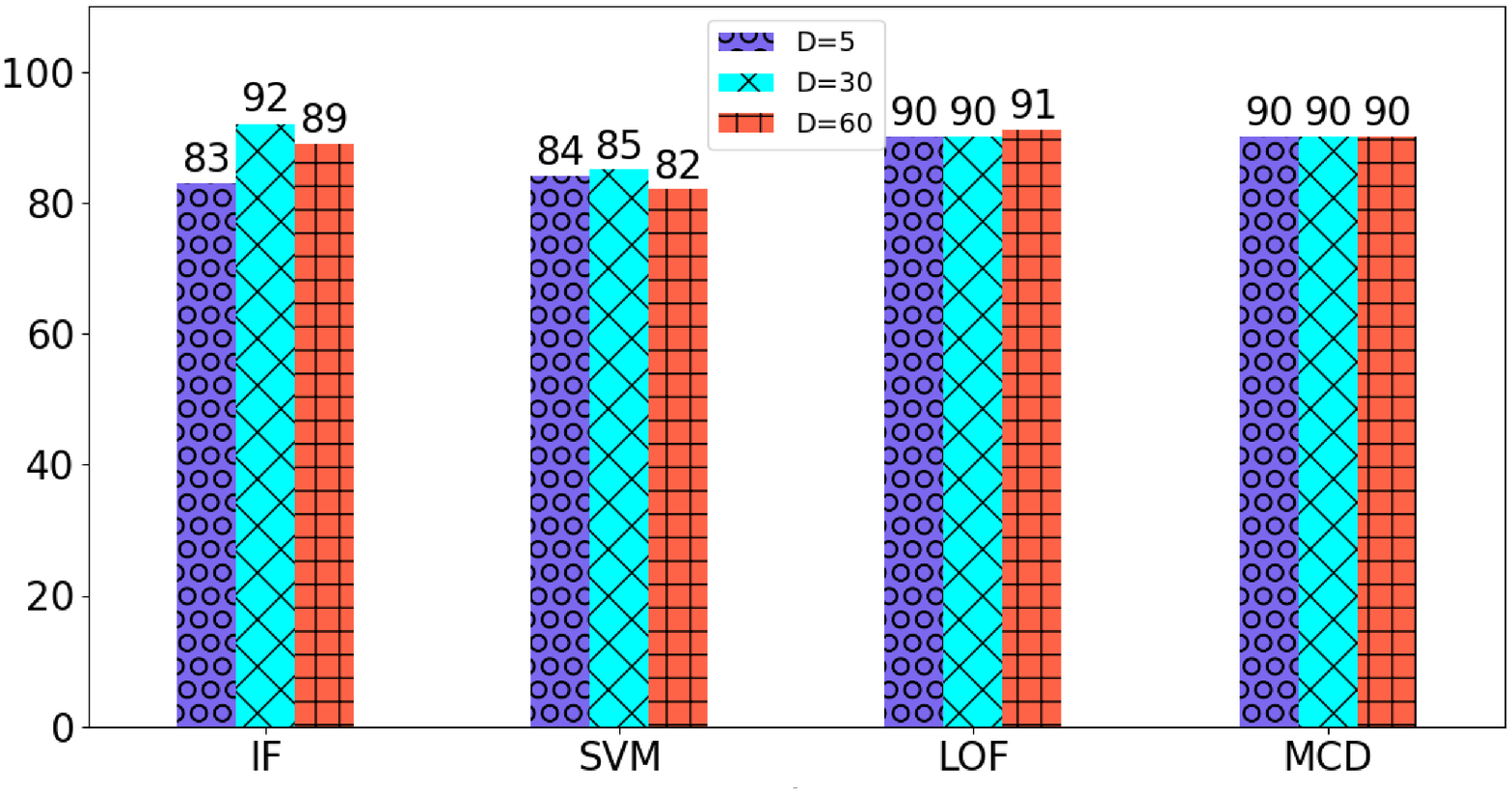}}
\subfloat[$20\%$ PUEA data]{
\includegraphics[width=0.48\columnwidth, height=1.5in]
{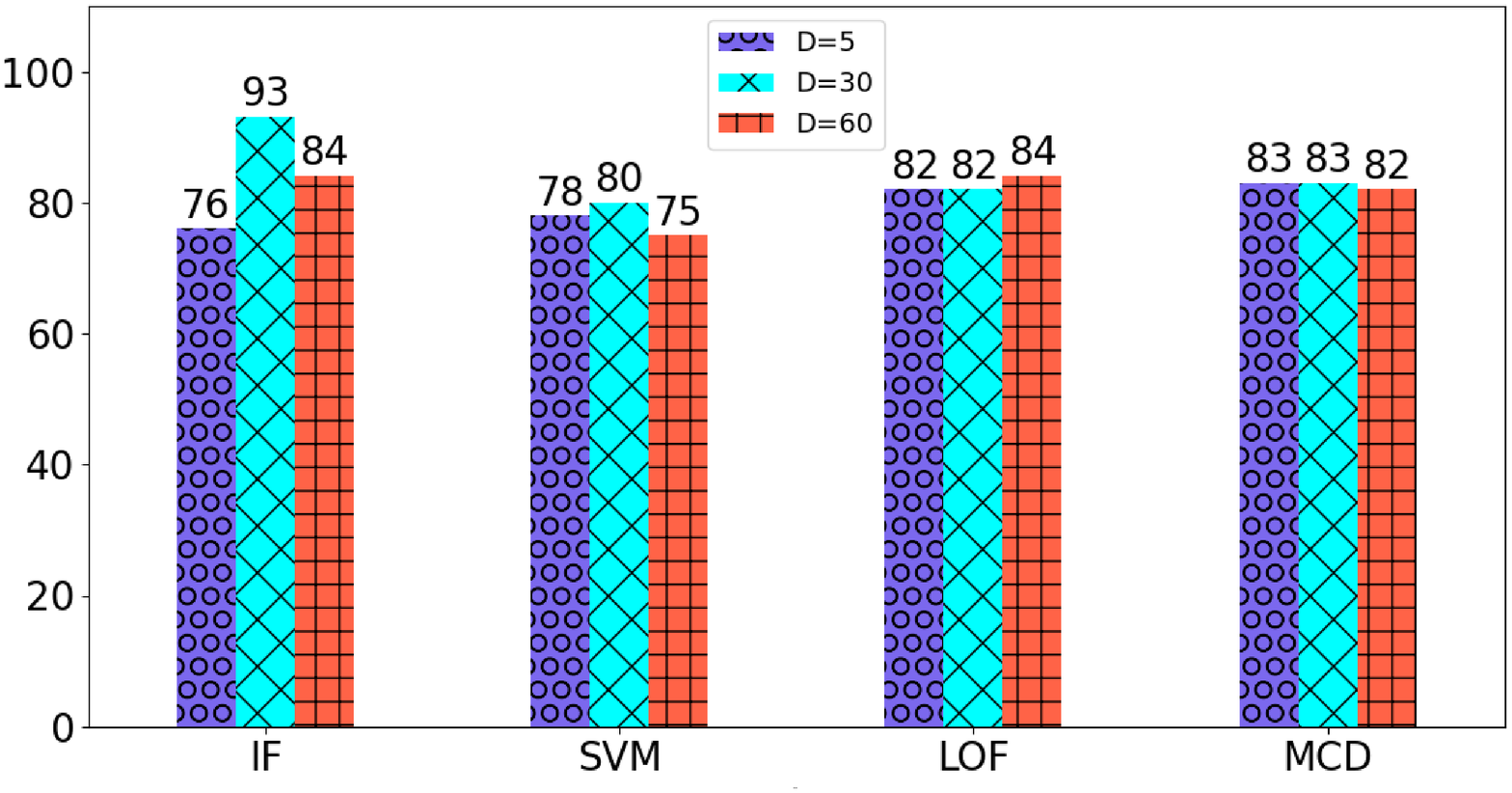}}
\caption{Accuracy: [Outside SU region]}
\label{figureAccuracyOutside}
\end{figure}

\begin{figure}[h]
\centering
\subfloat[$10\%$ PUEA data]{
\includegraphics[width=0.48\columnwidth, height=1.5in]
{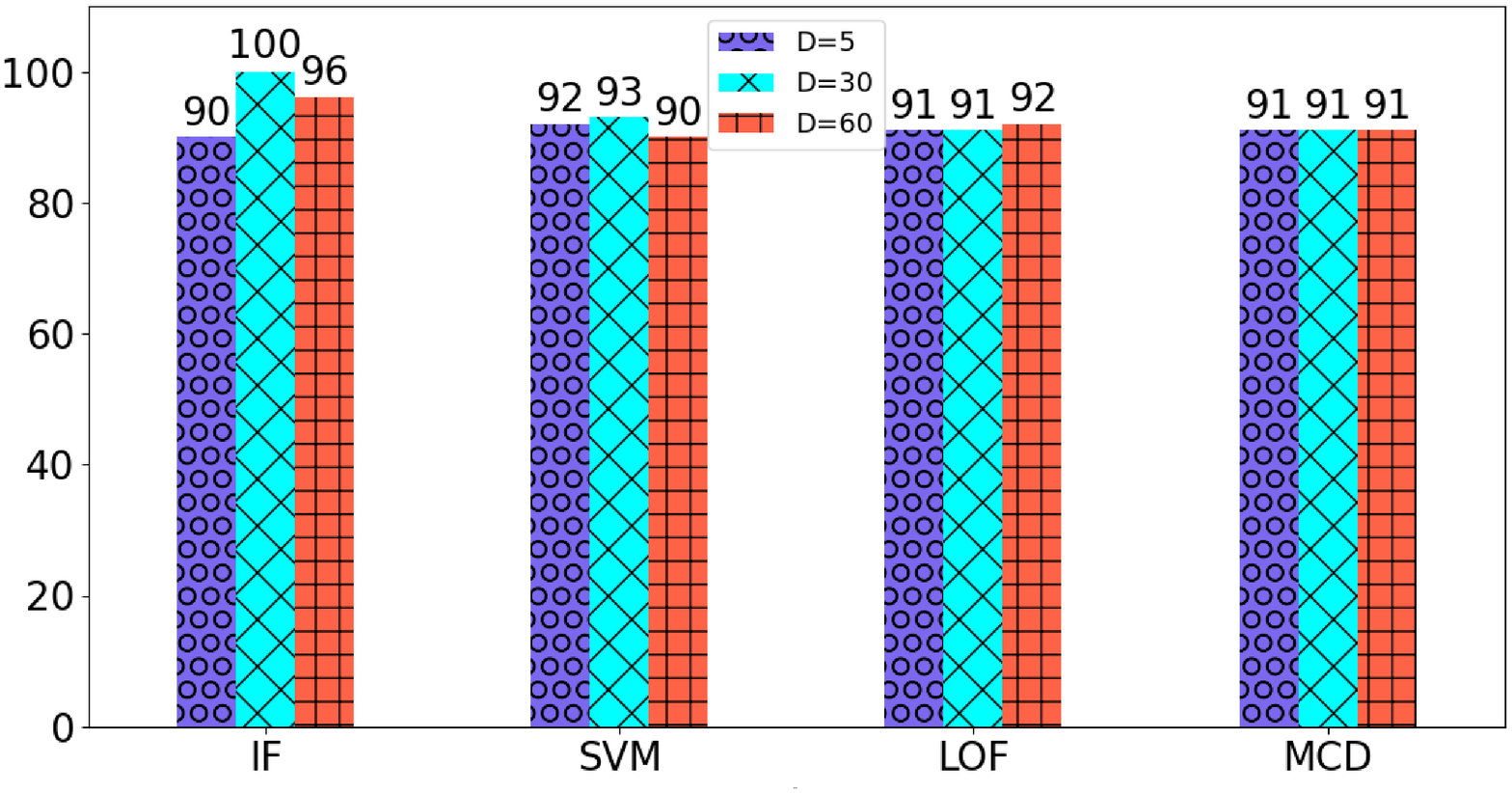}}
\subfloat[$20\%$ PUEA data]{
\includegraphics[width=0.48\columnwidth, height=1.5in]
{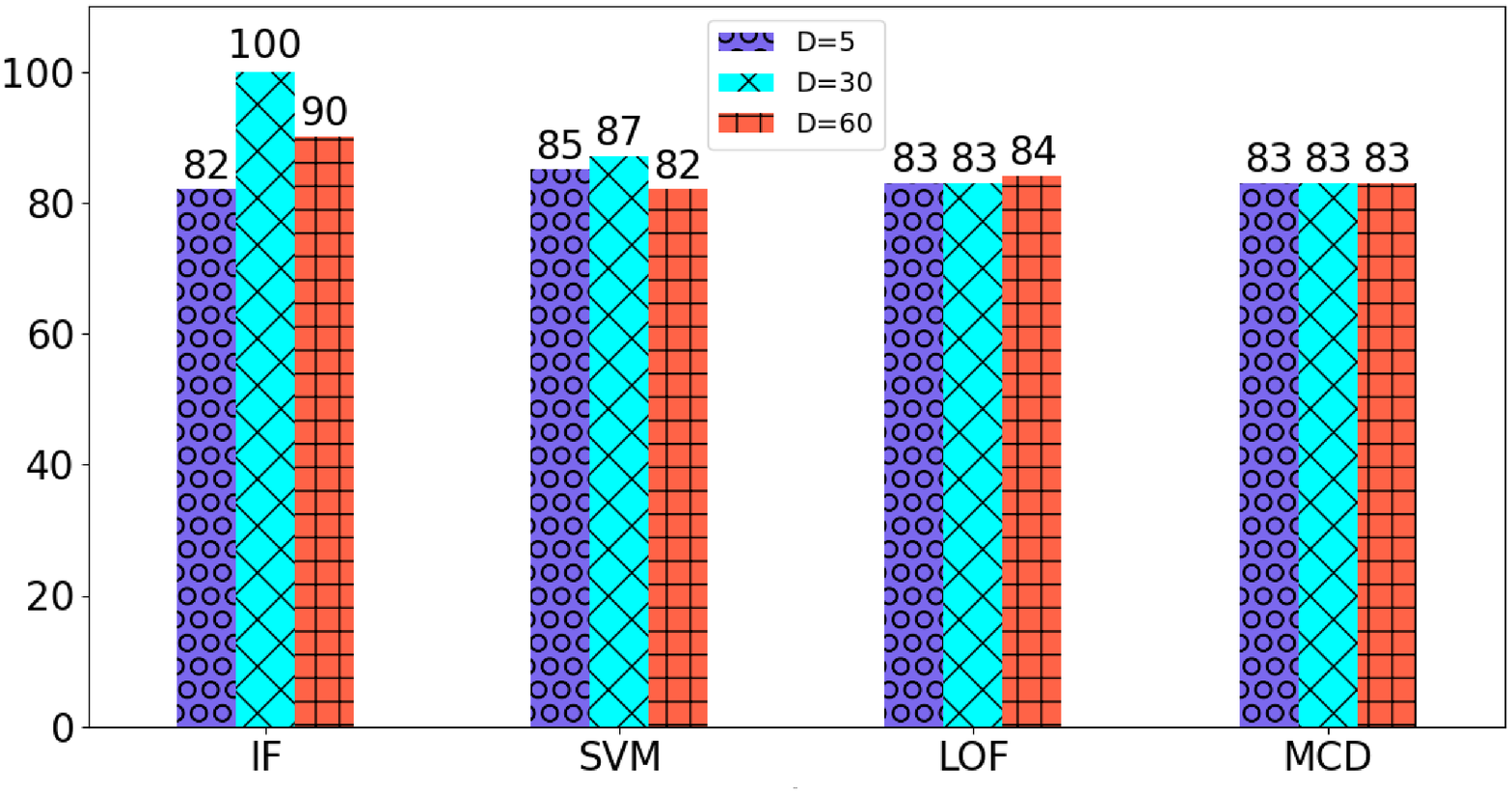}}
\caption{Precision: [Outside SU region]}
\label{figurePrecisionOutside}
\end{figure}

\begin{figure}[h]
\centering
\subfloat[$10\%$ PUEA data]{
\includegraphics[width=0.48\columnwidth, height=1.5in]
{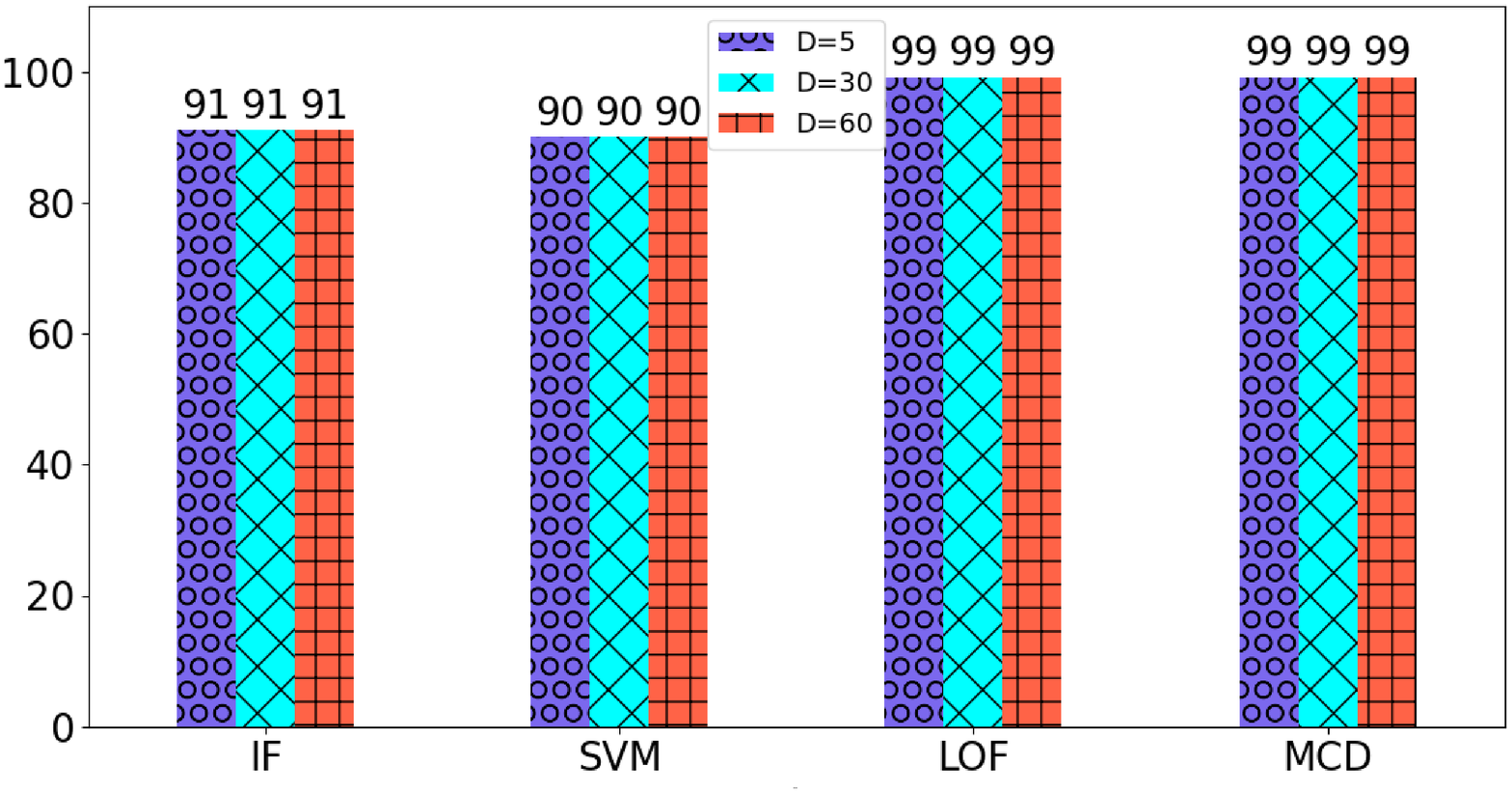}}
\subfloat[$20\%$ PUEA data]{
\includegraphics[width=0.48\columnwidth, height=1.5in]
{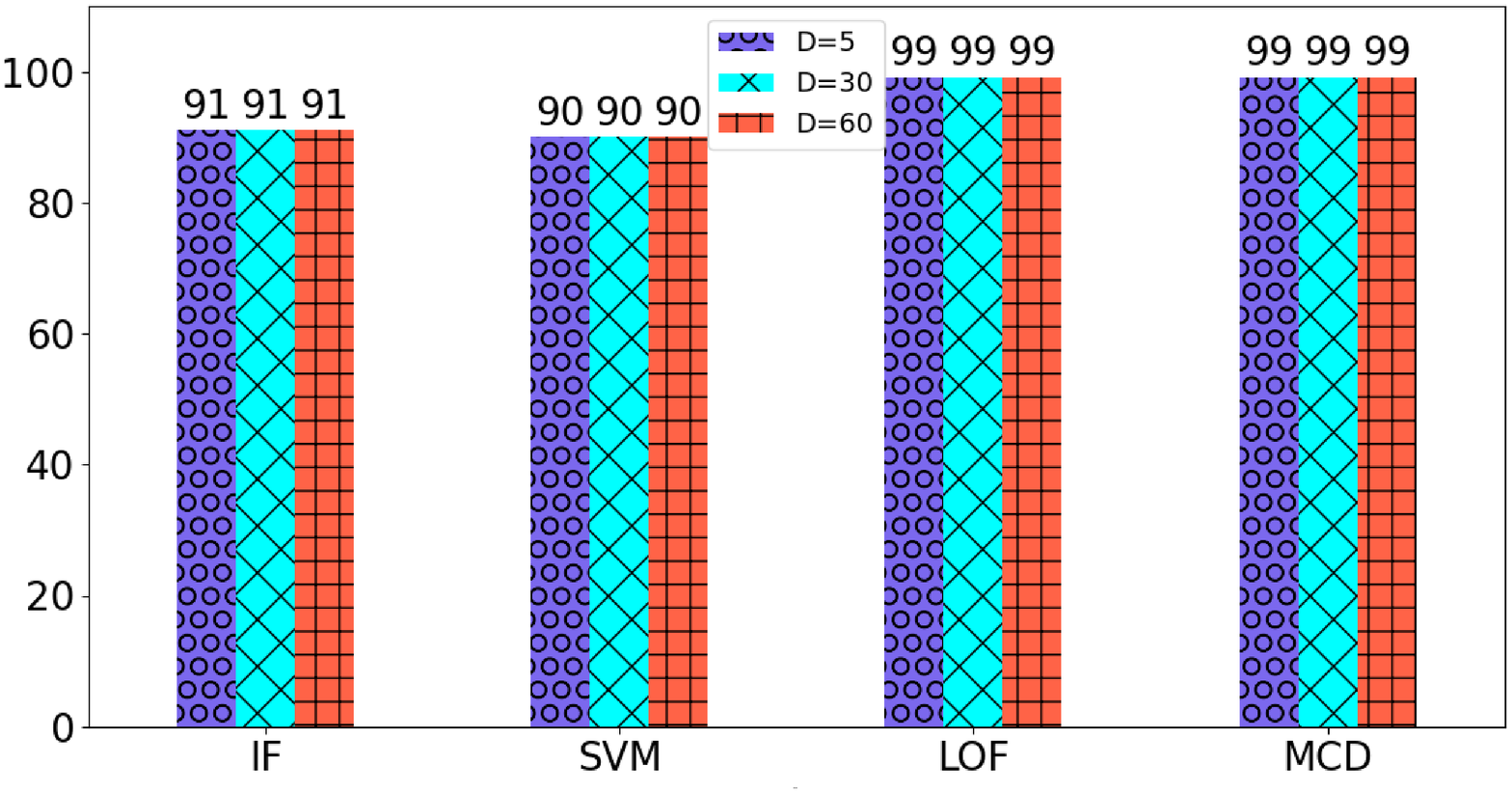}}
\caption{Recall: [Outside SU region]}
\label{figureRecallOutside}
\end{figure}

\begin{figure}[h]
\centering
\subfloat[$10\%$ PUEA data]{
\includegraphics[width=0.48\columnwidth, height=1.5in]
{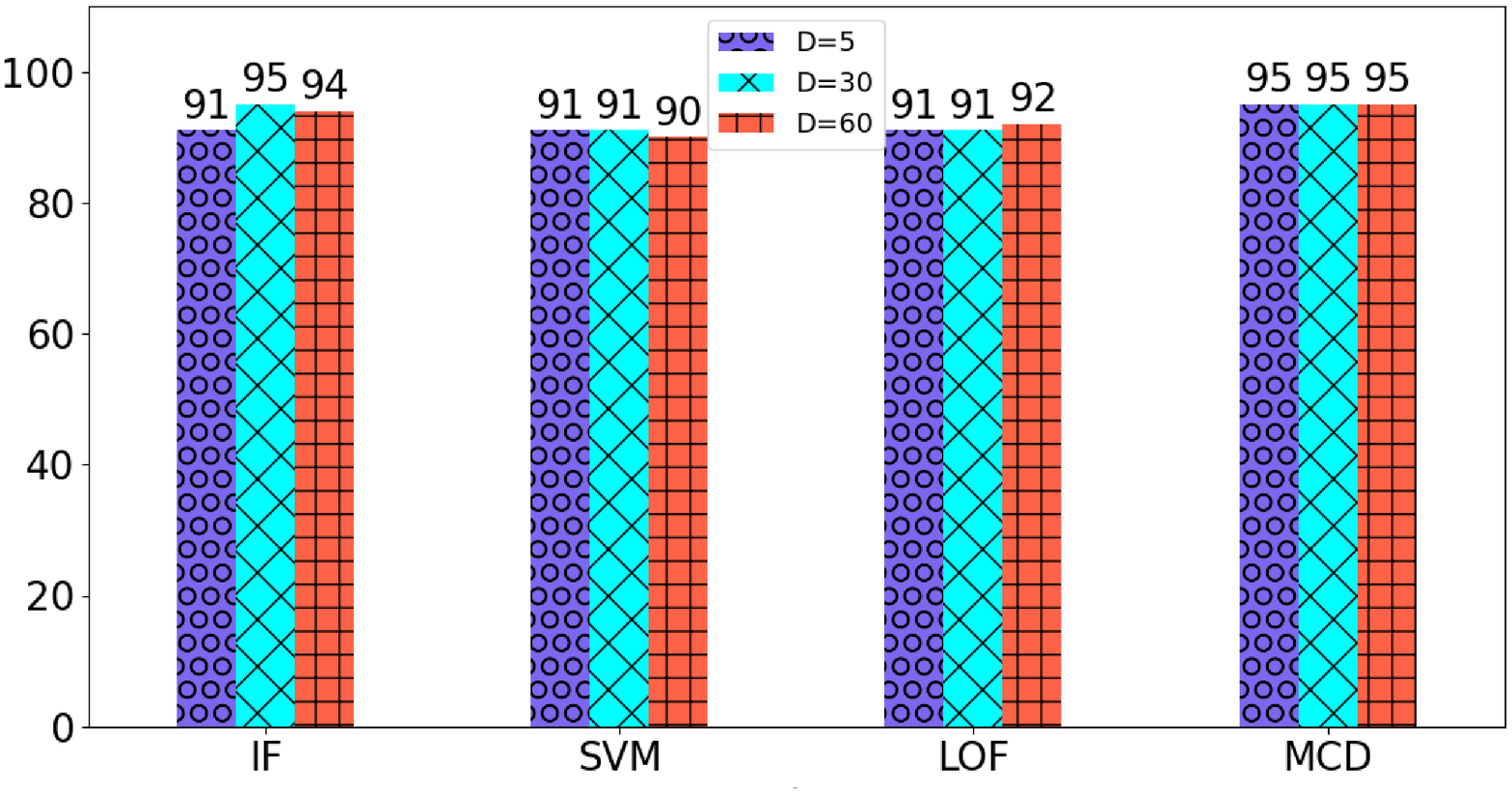}}
\subfloat[$20\%$ PUEA data]{
\includegraphics[width=0.48\columnwidth, height=1.5in]
{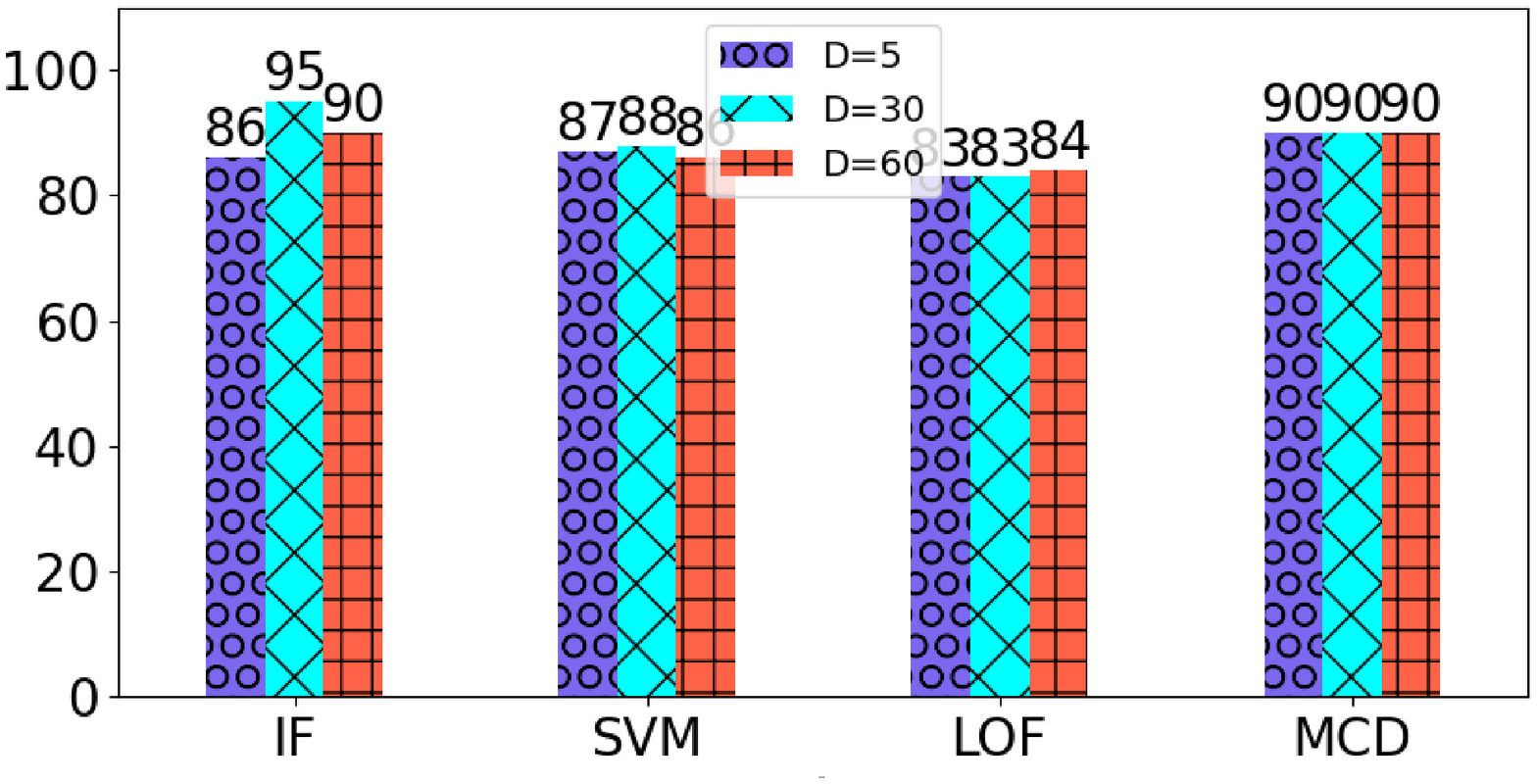}}
\caption{F1-score: [Outside SU region]}
\label{figureF1scoreOutside}
\end{figure}

Additionally, we employ $k$-fold cross validation 
on the two test data-sets. 
We partition the data-set into $k$ fixed-sized 
divisions of equal size. Then, we choose one partition as the test data and the 
rest $k-1$ 
partitions as the training data. We, then train the model using the training 
data and test using the test data. 
This process is repeated for $k$ iterations by choosing a 
distinct partition as the test data in each iteration. The performance results shown are determined by averaging over the $k$ iterations. Tables 
\ref{table:inside10} 
and 
\ref{table:inside20} 
give the accuracy 
for different values of $k$.   
Here, the value of $k$ does not
seem to have any significant effect on the accuracy.
\subsection{PU Outside SU Region}

The results presented in the previous subsection are for 
the case when the PU and the attacker are in the 
same region as the SUs 
(as illustrated in Fig. 
\ref
{figureSUregion}(a)). 
In this subsection, we present the results when they are 
outside the region 
(as illustrated in 
Fig. \ref{figureSUregion}(b)). 
This results are obtained so as to check whether there 
is any effect of varying distance and orientation 
between the PU and 
the SUs. 
%

The corresponding graphs that depict accuracy are given in 
Fig. 
\ref{figureAccuracyOutside}. We find that even if the PU 
and attacker 
are farther away from the SUs, the performance of the 
ML-based detection system is good. This can be seen also 
w.r.t. other performance metrics such as \emph{Precision}, \emph{Recall} and \emph{F1-score} as illustrated in 
Figs. 
\ref{figurePrecisionOutside} to 
\ref{figureF1scoreOutside}. 

\begin{table}[h!]
  \begin{center}    
    \begin{tabular}{|l|c|c|c|c|} 
\hline
\multicolumn{1}{|c|}{k} & IF & SVM & MCD & LOF\\
\multicolumn{1}{|c|}{ } & acc.(\%) & acc.(\%) & acc.(\%) &acc.(\%)\\
\hline
2 & 82 &84 &90&90\\
5& 82&84&89&90\\
10&82&84&89&90\\
20&82&84&90&90\\
\hline
   \end{tabular}
\caption{k-fold cross validation: 10\% PUEA data [Outside SU region]}
\label{table:outside10}
\end{center}
\end{table}
\begin{table}[h!]
  \begin{center}    
    \begin{tabular}{|l|c|c|c|c|} 
\hline
\multicolumn{1}{|c|}{k} & IF & SVM & MCD & LOF\\
\multicolumn{1}{|c|}{ } & acc.(\%) & acc.(\%) & acc.(\%) &acc.(\%)\\
\hline
2 & 75 &78 &82&82\\
5& 75&78&82&82\\
10&75&78&82&82\\
20&74&78&82&82\\
\hline
   \end{tabular}
\caption{k-fold cross validation: 20\% PUEA data [Outside SU region]}
\label{table:outside20}
\end{center}
\end{table}

For this scenario, we also apply $k$-fold cross validation. The corresponding accuracy results are 
given in Tables   
\ref{table:outside10}
and 
\ref{table:outside20} 
for different values of $k$. 
Here again, the value of $k$ does not
seem to have any significant effect on the accuracy.
\section{Conclusions} \label{conclusion}
This research studies the use of supervised machine 
learning, viz., one-class classification as a tool for 
detecting PUEA in infrastructure-based CRN. Classifiers such as Isolation Forest (IF), SVM, MCD and LOF are examined. 
We show through simulation how audit data of the sensing reports collected at the FC can be processed to 
characterize the behaviour 
of a PU transmission, thereby generating the training data-set. To  
the best of our knowledge, there has been no similar study that employs the rich sensed data collected at the FC for 
characterizing a PU transmission. Subsequently, we employ one-class classification algorithms and train 
them using the data-set thus created to 
build models that detect PUEA. 
Of all the techniques, SVM performs the worst with inconsistent performance. 
Isolation Forest (IF) performs better than SVM but not as good as MCD and LOF. MCD and LOF are the most robust, showing consistently very good performance. 




\end{document}